\documentclass[%
reprint,
groupedaddress,
amsmath,amssymb,
aps,
pra,
floatfix,
]{revtex4-1}

\usepackage{graphicx}
\usepackage{dcolumn}
\usepackage{bm}

\usepackage[mathlines]{lineno}

\usepackage{amsmath}
\usepackage{placeins}  
\usepackage{textcomp}
\usepackage{gensymb} 

\usepackage{mathtools}
\DeclarePairedDelimiter\bra{\langle}{\rvert}
\DeclarePairedDelimiter\ket{\lvert}{\rangle}
\DeclarePairedDelimiterX\braket[2]{\langle}{\rangle}{#1 \delimsize\vert #2}

\begin{document}

\preprint{APS/123-QED}

\title{Micromachined integrated quantum circuit containing a superconducting qubit}

\author{T. Brecht}
\email{teresa.brecht@yale.edu}
\author{Y. Chu}
\author{C. Axline}
\author{W. Pfaff} 
\author{J. Z. Blumoff} 
\author{K. Chou}

\author{L. Krayzman}
\author{L. Frunzio}
\author{R. J. Schoelkopf}
\affiliation{Department of Applied Physics, Yale University, New Haven, Connecticut 06511, USA}
\affiliation{Yale Quantum Institute, Yale University, New Haven, Connecticut 06520, USA}

\date{\today}
\begin{abstract}
We present a device demonstrating a lithographically patterned transmon integrated with a micromachined cavity resonator. 
Our two-cavity, one-qubit device is a multilayer microwave integrated quantum circuit (MMIQC), comprising a basic unit capable of performing circuit-QED (cQED) operations. 
We describe the qubit-cavity coupling mechanism of a specialized geometry using an electric field picture and a circuit model, and finally obtain specific system parameters using simulations. 
Fabrication of the MMIQC includes lithography, etching, and metallic bonding of silicon wafers. 
Superconducting wafer bonding is a critical capability that is demonstrated by a micromachined storage cavity lifetime $34.3~\mathrm{\mu s}$, corresponding to a quality factor of 2 million at single-photon energies. 
The transmon coherence times are $T_1=6.4~\mathrm{\mu s}$,  and $T_2^{Echo}= 11.7~\mathrm{\mu s}$. 
We measure qubit-cavity dispersive coupling with rate $\chi_{q\mu}/2\pi=-1.17~$MHz, constituting a Jaynes-Cummings system with an interaction strength $g/2\pi=49~$MHz. 
With these parameters we are able to demonstrate cQED operations in the strong dispersive regime with ease. 
Finally, we highlight several improvements and anticipated extensions of the technology to complex MMIQCs.
\end{abstract}

\maketitle


\section{Introduction}

Quantum circuits are soon reaching size and complexity that puts extreme demands on input/output connections as well as selective isolation among internal elements. Continued progress will require 3D integration and RF packaging techniques\cite{KatehiLPB:2001dl,QuantumSocket} that allow for scaling.  Indeed, there are numerous developed technologies waiting to see fruitful implementation in the field of circuit-QED (cQED), both from room temperature microwave devices\cite{brown_microwave_1999,Katehi:1997wc} and complex superconducting circuits\cite{SFQ_STolpygo,Tolpygo,Bintley:2012is}.
To address this opportunity and the associated challenges for quantum coherence, we recently proposed the multilayer microwave integrated quantum circuit (MMIQC) architecture\cite{Brecht:2016npj}, which adapts many existing circuit design and fabrication techniques to cQED.
A crucial step towards this vision is the demonstration of superconducting micromachined cavities\cite{Brecht:2015apl}, which can be used as quantum memories or as shielding enclosures to prevent cross-talk in more complex quantum computing devices. However, integrating transmons into these micromachined cavities has yet to be discussed, and is not a trivial matter of replicating the common methods in either existing planar or 3D cQED circuits. Fortunately, the flexibility and durability of MMIQC hardware affords many possibilities for qubit integration.

In this work, we demonstrate one such possiblity through the design, fabrication, and characterization of a quantum device containing a transmon qubit coupled to a superconducting micromachined cavity. It forms a simple MMIQC capable of performing cQED operations. The techniques shown here can be improved and extended to realize more complex quantum circuitry. 

\begin{figure}
\centering
\linespread{1}
\includegraphics[scale = 1,angle=0]{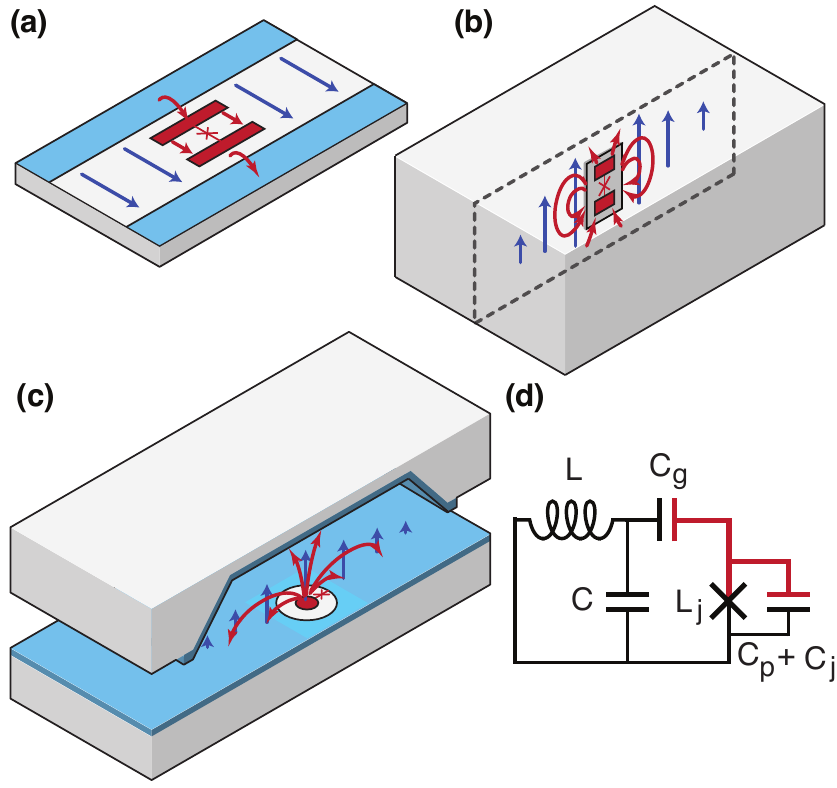}
\caption{
 	Illustrations of dipole coupling in cQED. Electric dipole moment orientations for typical transmons are fabricated to align with (a) the electric field of a planar transmission line resonator or (b) the electric field of an encapsulating 3D cavity. Blue arrows show electric field lines of the each resonator's fundamental mode, and red arrows show electric field lines of the transmon mode.
    (c) The aperture transmon fields couple to the fundamental mode of the micromachined cavity in the device discussed in this work. For clarity, the diagram shows an exploded cross-sectional view of two substrate wafers, and it is not to scale.
    (d) Schematic circuit diagram of the aperture transmon and electromagnetic resonator. The red coloring corresponds to the central island.
    }
\label{Fig:coupling}
\end{figure}

\section{Qubit-resonator coupling}

Coupling between an electromagnetic resonator and a transmon occurs via shared electric and magnetic fields of their respective modes. 
Both planar circuits and 3D qubits use a simple dipole antenna structure aligned with the electric field of a transmission line, waveguide or cavity.\cite{Paik:2011hd} 
These common schemes are diagrammed in Fig.~\ref{Fig:coupling}(a)-(b). 
It would be impractical to use a similar scheme for coupling qubits to micromachined cavities because of the extreme aspect ratio imposed by the wafer height. 
Instead, we desire to achieve the same coupling while limiting ourselves to planar fabrication and wafer stacking. 
A circuit can be patterned on one of the cavity walls such that the electromagnetic fields couple to those of the cavity, as shown in Fig.~\ref{Fig:coupling}(c).  

The coupling of an ``aperture transmon'' to a resonator using fields that are out of the plane of transmon fabrication is described in Ref.~\cite{Minev:2016tl}. In the device of the present work, the coupling can be understood by analyzing the overlap between the electric fields of the transmon mode and those of the adjacent cavity mode(s), and also by an equivalent circuit model. Translation of the aperture transmon away from center of the cavity wall results in a mixture of electric (charge accumulation) and magnetic (current flow) coupling. However, the aperture transmon's central location maximizes total coupling.

The schematic circuit diagram is depicted in Fig.~\ref{Fig:coupling}(d). A single Josephson junction connects the central island to the rest of the cavity wall. It is accompanied by a junction capacitance ($C_j$),which is small compared to the other capacitors in the system: First, there is a capacitance across the open annulus between the island and the rest of the lower cavity wall ($C_p$). Second, there is a capacitance across the gap between the island and the opposite wall of the cavity ($C_g$). Lastly, there is capacitance $C$ associated with the walls of the cavity, which combines with an effective inductance $L$ to create the LC-resonator characteristic of the cavity's fundamental mode at frequency $\omega_{\mu}/2\pi$.

The system of qubit excitations and resonator photons displays a Jaynes-Cummings interaction: $ \hbar g (a^{\dagger} \sigma^- + a \sigma^+ ) $, where $a^{\dagger} ~ (a) $ creates (annihilates) a photon and $\sigma^+ ~ (\sigma^-) $ creates (annihilates) an excitation of the qubit.
The coupling rate $g=e V_0 \beta / \hbar$ is a function of the capacitances, $\beta = C_g/(C_g+C_p+C_j),$
and $V_0 = \sqrt{\hbar \omega_{\mu} / 2 C}$. This circuit picture using capacitances reveals the relationships between device geometry and coupling strength, analogous to other cQED hardware designs.\cite{Blais:2004kn}  See supplementary material for details.\cite{SUPPLEMENT} In this work, we operate such a system in the the strong dispersive limit, where the frequency detuning between resonator and qubit is much greater than the interaction rate ($|\Delta| \gg g$) and the interaction rate is much greater than the decay rates of the qubit or cavity ($g \gg \gamma, \kappa$).

In this strong dispersive limit, we approximate the applicable Hamiltonian as

\begin{multline}  \label{hamiltonian}
\frac{H}{\hbar} = 
\sum_{i}\omega_i a_i^{\dagger} a_i 
- \sum_{i \neq j}\chi_{ij} a_i^{\dagger} a_i a_j^{\dagger} a_j
 - \sum_{i} \frac{\alpha_i}{2} a_i^{\dagger 2} a_i^2
\end{multline}
including an arbitrary number of modes. Each mode has transition frequency $\omega_i$ between its first two levels and an anharmonicity $\alpha_i$, which is greatest for the transmon. Each pair of modes interacts via a dispersive shift of strength $\chi_{ij}$.

\section{Device Implementation}

\begin{figure*}[ht]
\centering
\linespread{1}
\includegraphics[scale = 1,angle=0]{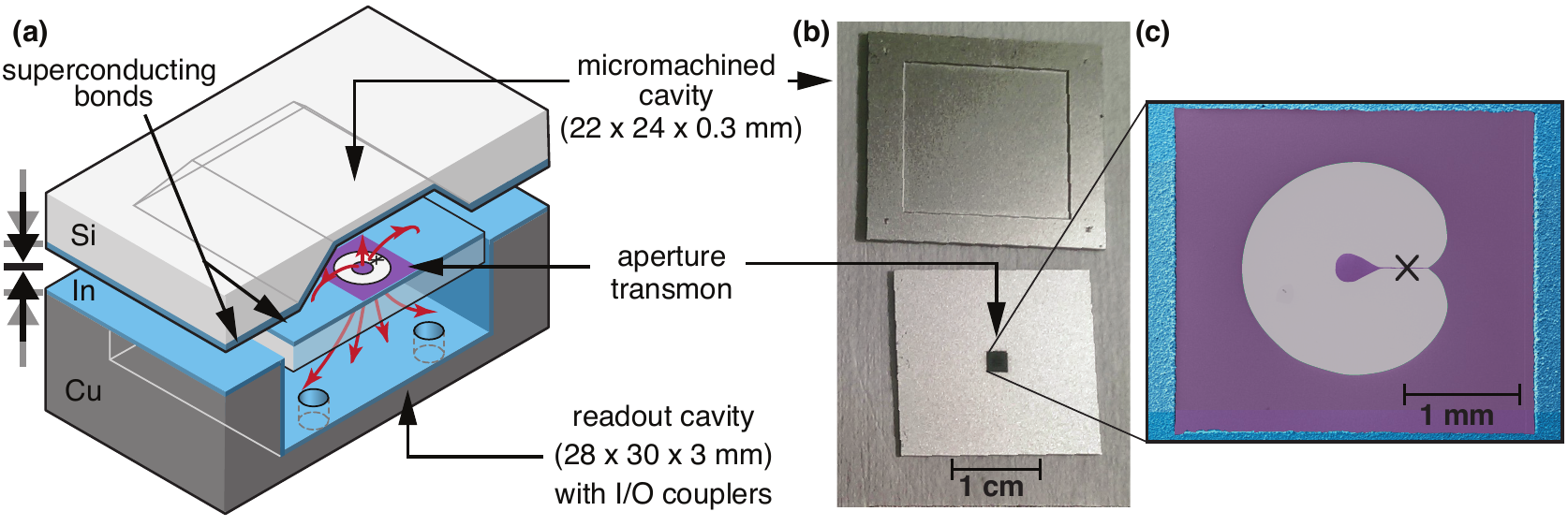}
\caption{
	(a) Sketch of device. For clarity, the image shows an exploded cross-sectional view that is not to scale. The annular structure has electric dipole moment components in two opposing directions, both perpendicular to the plane of fabrication. Red arrows show electric field lines of the transmon mode, and the transmon chip is shown semi-transparent.
    (b) Photograph of the micromachined cavity chip (top) and transmon chip (bottom).
 	(c) False colored SEM image of the aperture transmon, with silicon in grey, aluminum in purple, and indium in blue. The shape of the electrodes is described in the supplementary material\cite{SUPPLEMENT}. An `X' indicates the Josephson Junction position, interrupting a 50~$\mu$m wide lead connecting the inner island to the remainder of the cavity wall.
    }
\label{Fig:device}
\end{figure*}

The first MMIQC prototype is designed to have coherent quantum modes that have sufficient coupling rates between them, and allow for manipulation and measurement with microwave pulses. The device featured in this work consists of three quantum objects. An aperture transmon couples simultaneously to two cavity modes. In addition to a micromachined cavity, a second cavity made by traditional metal machining is incorporated to compose a two-cavity/one-qubit MMIQC device. This cavity allows readout of the transmon and micromachined cavity states through two pins leading to coaxial cables for microwave access to the system, and is hereafter referred to as the ``readout cavity''. The device, shown in Fig.~\ref{Fig:device}, displays a hybrid multilayer construction, including silicon wafers and conventionally machined metals united by indium bonding on flat surfaces. 
The integration of the machined 3D cavity demonstrates the aperture transmon's bipartite coupling and provides a convenient way of connectorizing the MMIQC.


Next, we must choose design parameters to realize the MMIQC. We also impose that the qubit is in the transmon regime, with suppressed charge dispersion\cite{Koch:2007,Schreier:2008}. The shape and position of the aperture transmon affect properties of the system between which tradeoffs are considered. For example, the size of the inner island must be large enough to create a measurable $g$ by the capacitance contribution $C_g$. However, if the inner island is too large, the anharmonicity is reduced, limiting speed of manipulation pulses.
Scaling trends of $g$ changing with respect to several relevant geometrical parameters are included in supplementary material\cite{SUPPLEMENT}. 
The coupling between the qubit and the micromachined cavity, $\chi_{q\mu}$, and that of the qubit and the readout cavity, $\chi_{qr}$, are also adjusted by choice of heights of each cavity and thickness of qubit substrate. We perform simulations in order to confirm our understanding of the qubit-cavity coupling and to aid geometry optimization more precisely. We model the entire system using a full 3D electromagnetic simulation using a finite element solver followed by blackbox quantization analysis\cite{Nigg:2012jj}. For the design featured in this work, the anharmonicity is designed to be $\alpha_q= -E_c = -204~$MHz, and the Josephson energy is $E_J/h = \Phi_0^2 L_J / 2 \pi h = 39~$GHz. ($E_J/E_C = 193$.)

We now briefly describe how the device is constructed. The multilayer device is fabricated as three separate parts (see Fig.~\ref{Fig:device}(a)) that are finally bonded together with a metal that superconducts. 
The micromachined cavity chip is created by wet etching a rectangular pit to depth of $300~\mathrm{\mu m}$ in silicon, followed by metalization with $10~\mathrm{\mu m}$ indium.\cite{Brecht:2015apl,SUPPLEMENT}
The transmon chip requires three metalization steps on a $325~\mathrm{\mu m}$ thick silicon wafer:  a patterning of gold by liftoff, ebeam lithography and shadow angle evaporation of the aluminum Josephson junction, and masking this junction before electroplating $10~\mathrm{\mu m}$ of indium onto the gold.
The readout cavity is milled out of OFHC copper and electroplated with $30~\mathrm{\mu m}$ indium.
The three components are bonded together between parallel plates at 120\degree~C in two steps.
Once assembled, coaxial pin couplers are added to the readout cavity and device is thermally anchored to the baseplate of a dilution refrigerator reaching a base temperature of 15~mK.
See supplementary material\cite{SUPPLEMENT} for additional fabrication and bonding details.

\begin{table}[h]
\caption{
Measured device parameters. The cross-Kerr interaction with the qubit mode is denoted $\chi_q$, and anharmonicity is $\alpha$. Simulated parameters are in italics, and all other parameters are measured except the anharmonicities of the cavities, which are calculated by $ \alpha = \chi_{q}^2/4\alpha_q$.\cite{Nigg:2012jj}
}
\label{mytable}

\bgroup
\def\arraystretch{1.25}
\begin{tabular}{c|ccc}
							& Readout&		 & $\mu$-Machined \\
Mode                		& cavity & Transmon &  cavity 		\\ \hline \hline
Frequency (MHz)     		& 6973.4 & 7351.4 & 9377.2   \\
$\left[ simulated \right]$	&\it{6945.1} & \it{7322.0}	& \it{9258.0}\\ \hline
$\alpha_i/2\pi$ (MHz) 		& -0.012  & -209.8   & -0.002   \\ 
$\left[ simulated \right]$	& \it{-0.004}& \it{-204.3} & \it{-0.002}\\ \hline
$\chi_{qi}/2\pi$ (MHz) 		& -3.84   	 &  -       & -1.17     \\
$\left[ simulated \right]$	& \it{-3.22} &	 -	     & \it{-1.25}\\ \hline
$\chi_{ri}/2\pi$ (MHz) 		& -          &   -       & -0.020     \\
$\left[ simulated \right]$	& -			 &	 -	     & \it{-0.004}\\ \hline
$T_1$ ($\mu$s)     			& 1.0     & 6.4      &  34.3    \\ \hline
$T_2^{R}$ ($\mu$s)    		&   -   &  9.5   & - \\ 
$T_2^{Echo}$ ($\mu$s) 		&   -   & 11.7   & - \\ \hline
\end{tabular}
\egroup
\end{table}

\begin{figure*}[ht]
\centering
\linespread{1}
\includegraphics[scale = 1,angle=0]{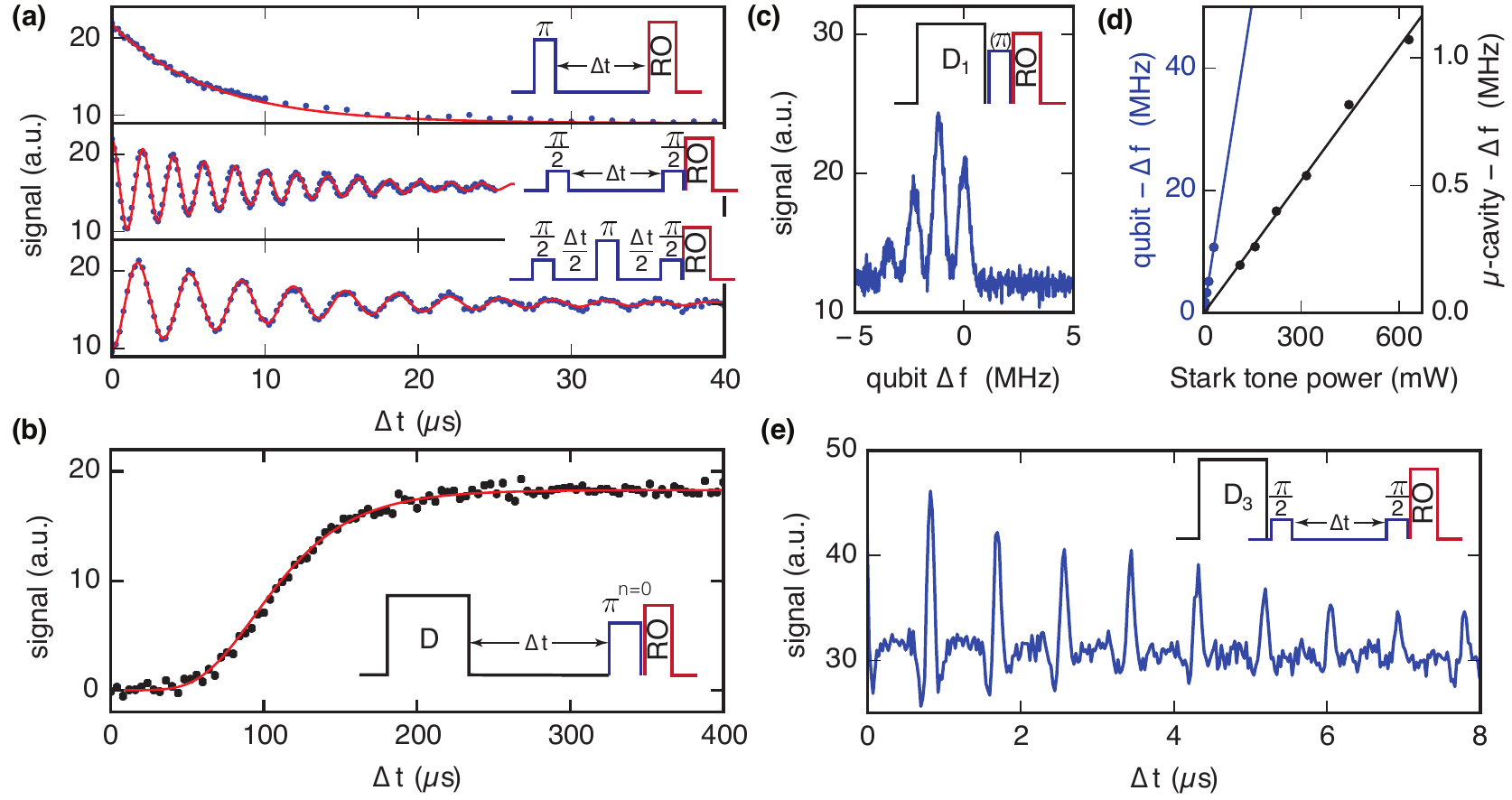}
\caption{
    (a) Qubit energy relaxation is fit to a single exponential (red line) with $T_1= 6.4~\mu$s. Ramsey dephasing time $T_2^R = 9.5~\mu$s, measured here using 400 kHz detuning from the qubit frequency. Using a Hahn echo sequence, we find $T_2^{echo} = 11.7~\mu$s, measured here using 300 kHz detuning from the qubit frequency.
    (b) Energy decay of the micromachined cavity is measured by applying a large displacement to this cavity, followed by a variable delay, followed by a spectrally narrow selective $\pi$-rotation of the qubit conditioned on there being no photons in the readout cavity ($n=0$). Using a Poissonian decay fit (red line), we find $T_1=34.7~\mu$s. At 9.4 GHz, this decay time corresponds to quality factor $Q = 2$ million.
    (c) We observe number splitting of the qubit in spectroscopy after displacing the micromachined cavity by one photon. The spacing between the peaks indicates $\chi_{q\mu}/2\pi=1.17~$MHz 
    (d) A tone detuned 3 MHz above the readout cavity induces a Stark shift that affects both the qubit and micromachined cavity frequencies. We use the ratio of these slopes $\chi_{qr}/\chi_{r\mu}$ to determine $\chi_{r\mu}$.
    (e) In Ramsey interferometery following a displacement of the micromachined (storage) cavity, we observe revivals of the qubit state occurring at integer multiples of $2\pi/\chi_{q\mu} = 0.855~\mu$s.\cite{Vlastakis:2013ju}
    }
\label{Fig:ChiGraphs}
\label{Fig:T1T2Graphs}
\end{figure*}

\section{Experimental Results}

Successful cQED operation in this new hardware is demonstrated with measurements of coherence times and interactions between each quantum object of the MMIQC. Measurements of the relevant coherence times in the device are shown in Fig.~\ref{Fig:T1T2Graphs}(a)-(b) and summarized in Table \ref{mytable}. The qubit $T_1=6.4~\mu$s is on the order of other 3D transmons recently produced on silicon substrate in the same facility with similar methods\cite{Chu:2016} and $T_2^{Echo} \approx 1.8T_1$. The thermal population of the qubit excited state is $<3$ percent. The micromachined cavity has a lifetime of $34.7~\mu s$, which corresponds to a total quality factor $Q = 2$ million at single-photon energies.

Next, we find interaction strengths sufficiently large in relation to these coherences by showcasing some standard cQED functions. These measurements are shown in Fig.~\ref{Fig:ChiGraphs}(c)-(e). The dispersive coupling rate of the qubit to the readout cavity is $\chi_{qr}/2\pi=-3.84~$MHz, corresponding to interaction strength $g/2\pi=38~$MHz. In spectroscopy, we observe both resolved photon number splitting of the qubit (Fig.~\ref{Fig:ChiGraphs}(a)) and a qubit-state dependent shift of the micromachined cavity from a dispersive coupling rate $\chi_{q\mu}/2\pi=-1.17~$MHz. At detuning of $(\omega_q-\omega_\mu)/2\pi = -2.03~$GHz, this corresponds to $g/2\pi=49~$MHz. 

A final measurable parameter is the cross-Kerr interaction between the two cavities. The cavity cross-Kerr $\chi_{r\mu}$ is measured by relative comparison of $\chi_{qr}$ and $\chi_{r\mu}$.\cite{Leghtas853} A microwave pulse detuned 3 MHz above the readout cavity induces a Stark shift, which we use to precede single-side-band spectroscopy of both the qubit and micromachined cavity peaks. Both shift downward in frequency with increasing power of the Stark pulse. The slopes of this response are proportional to $\chi_{qr}$ and $\chi_{r\mu}$ respectively (Fig.~\ref{Fig:ChiGraphs}(b)).  We have independently determined $\chi_{qr}=-3.84~$MHz by readout cavity spectroscopy with and without a preceding qubit $\pi$-pulse. Finally, we find cross-Kerr $\chi_{r\mu}/2\pi=-20~$kHz, compared to a simulated value of $-4.4~$kHz.

As a further demonstration of the micromachined cavity's utility as a quantum memory, we perform Ramsey interferometry following a displacement that initializes the micromachined cavity to a coherent state of $\ket{\beta}$ with an average of three photons.
(Fig.~\ref{Fig:ChiGraphs}(c)) In this experiment, we prepare an initial state $\ket{\beta}_{\mu} \otimes \{ \ket{g} + \ket{e} \}$, which precesses according to $e^{i \chi_{q\mu} t a^{\dagger} a \ket{e}\bra{e}}$.\cite{Vlastakis:2013ju}  Qubit state revivals occur at time intervals $2\pi / \chi_{q\mu}$, consistent with our spectroscopic measurements of $\chi_{q\mu}$.

\section{Discussion of loss mechanisms}

We assess several potential loss mechanisms that could be limiting the coherence times in our device. 
All quantum circuits are subject to sources of loss associated with packaging and assembly that become more severe as complexity increases.\cite{Brecht:2016npj}
For example, loss occurs at seams where there is finite conductance, $g_{\mathrm{seam}}$, and non-zero admittance to surface currents, $y_{\mathrm{seam}}^{i}$, which may limit a mode $i$'s coherence time to $T_1 = g_{\mathrm{seam}}/y_{\mathrm{seam}}^{i}\omega_i$. 
In the multilayer architecture, these seams are present in the bonds between layers and interfaces between different materials.

In this device, there are two types of seams that could contribute to loss. 
The first consists of In/In bonds at the perimeter of the cavities. 
Using simulated surface currents, we calculate the admittance in the micromachined cavity mode to be $y_{\mathrm{In/In}}^{\mu}=16.0~/\Omega$m.  For the qubit mode, $y_{\mathrm{In/In}}^{q}=0.02~/\Omega$m, which is smaller becuase the surface currents are localized away from the In/In bond. Using the technique developed in \cite{Brecht:2015apl}, we are able to achieve an In/In bond conductance in our devices of $g_{\mathrm{In/In}} \approx 10^8~/\Omega$m.
If this were the only source of loss, it would limit the micromachined cavity to lifetime $100~\mu$s.
The second type of seam is a Al/Au/In transition in a $3\times3~$mm square shape around the Al aperture transmon region. 
Using simulated surface currents, we calculate the admittances in the micromachined cavity and qubit mode:
$y_{\mathrm{Al/Au/In}}^{\mu}=0.17~/\Omega$m and 
$y_{\mathrm{Al/Au/In}}^{q}=0.52~/\Omega$m. 
Independent measurements of stripline resonators fabricated with like procedures show that $g_{\mathrm{Al/Au/In}} \approx 4.2\times10^5~/\Omega$m.\cite{SUPPLEMENT} The resulting limitation on the qubit mode lifetime is $T_1 < g_{\mathrm{seam}}/y_{\mathrm{seam}}^{q}\omega_q \approx 20~\mu$s. Limitation on the cavity mode lifetime due to this seam is $T_1 \approx 40~\mu$s.

We also verified that qubit and micromachined cavity lifetimes are not limited by the Purcell effect due to the overcoupled readout cavity. We simulated that the upper bounds to the qubit and micromachined cavity lifetimes due to this effect are 200 and 500~$\mu$s respectively. In design, the Purcell limit of the micromachined cavity is mitigated by minimizing the area of the annular opening created by the aperture transmon between cavities. 

Also present here are surface dielectric and conductor loss mechanisms that are broadly studied in superconducting circuits. The particular shapes of the electrodes in Fig.~\ref{Fig:device}(c) are designed to minimize dielectric loss near the surface of the electrodes. The aperture transmon has smooth edges that are easily parameterized for optimization that includes consideration of surface participation ratios. See supplementary material\cite{SUPPLEMENT}.

\section{Outlook}
The device presented here demonstrates the integration of a transmon with a superconducting micromachined cavity, forming the first actual MMIQC. The coherence times and coupling rates are in the strong dispersive regime of cQED, enabling many quantum manipulations that are the precursor to large scale quantum information processing.

We remark that this achievement was made without extensive fabrication optimization using industrial scale tools, indicating process robustness and potential for improvement. For example, it is expected that surface cleaning will improve $g_{\mathrm{Al/Au/In}}$, a seam conductance relevant to both the qubit and cavity modes. Alternative MMIQC designs are being developed that contain different seams and minimize the use of normal metals like gold. Extended qubit lifetimes can be achieved by removal of silicon substrate in the junction area and more directed surface cleaning.\cite{Chu:2016} Furthermore, a wide range of coupling rates can be accessed by geometry modifications, some of which would require precise alignment and leveling control during wafer bonding. 

We have shown a proof-of-principle MMIQC that demonstrates the engineering of qubit-cavity coupling. There are numerous possible next steps using the design strategies and fabrication tools described in this work. For instance, the micromachined cavity and qubit can be addressed using microstrips on the side of the wafer opposite to the micromachined cavity wall and qubit fabrication, eliminating the machined readout cavity. The microstrips would be positioned above apertures in the metal of the cavity wall sized to create desired coupling rates. They can compose planar readout resonators and incorporate Purcell filtering\cite{Reed:2010bx}. As a further example, the addition of a second junction and flux-bias-line would constitute a frequency tunable device inspired by the concentric transmon in Ref. \cite{Braumuller:2016ig}. More sophisticated on-chip input/output circuitry, such as quantum limited amplifiers\cite{Castellanos,Bergeal:2010iu,Macklin307}, circulators\cite{Sliwa:2015cc,Kerckhoff2015}, and switching elements\cite{switch, Chapman2016}, will also be required for practical quantum information processing.
We anticipate that the techniques demonstrated here can be successfully employed toward integrating these elements into increasingly complex MMIQCs.

\section{Acknowledgments}

We thank Zlatko Minev, Kyle Serniak, Ioan Pop, Michel H. Devoret, Hanhee Paik and David Pappas for useful conversations, Subhajyoti Chaudhuri for assistance with simulations, Jan Schroers, and Emily Kinser for assistance with wafer bonding. 
This research was supported by the U.S. Army Research Office under grant W911NF-14-1-0011.
W.P. was supported by NSF grant PHY1309996 and by a fellowship instituted with a Max Planck Research Award from the Alexander von Humboldt Foundation. 
C.A. acknowledges support from the NSF Graduate Research Fellowship under Grant No. DGE-1122492.
Facilities use was supported by the Yale SEAS cleanroom, YINQE, and NSF MRSEC DMR-1119826.




%

\clearpage







\onecolumngrid

\begin{center}
\textbf{\large Supplemental Material: \\  \vspace{5mm} 
Micromachined integrated quantum circuit containing a superconducting qubit
}
\vspace{7mm}
\end{center}

\setcounter{equation}{0}
\setcounter{figure}{0}
\setcounter{table}{0}
\setcounter{page}{1}
\setcounter{section}{0}
\makeatletter
	\@addtoreset{figure}{section}
\renewcommand{\theequation}{S\arabic{equation}}
\renewcommand{\thefigure}{S\arabic{figure}}
\renewcommand{\thetable}{S\arabic{table}}
\renewcommand{\bibnumfmt}[1]{[S#1]}
\renewcommand{\citenumfont}[1]{S#1}


\section{Coupling trends}


\begin{figure*}[b]
\begin{minipage}{\textwidth}
\linespread{1}
\centering
\includegraphics[scale = 1,angle=0]{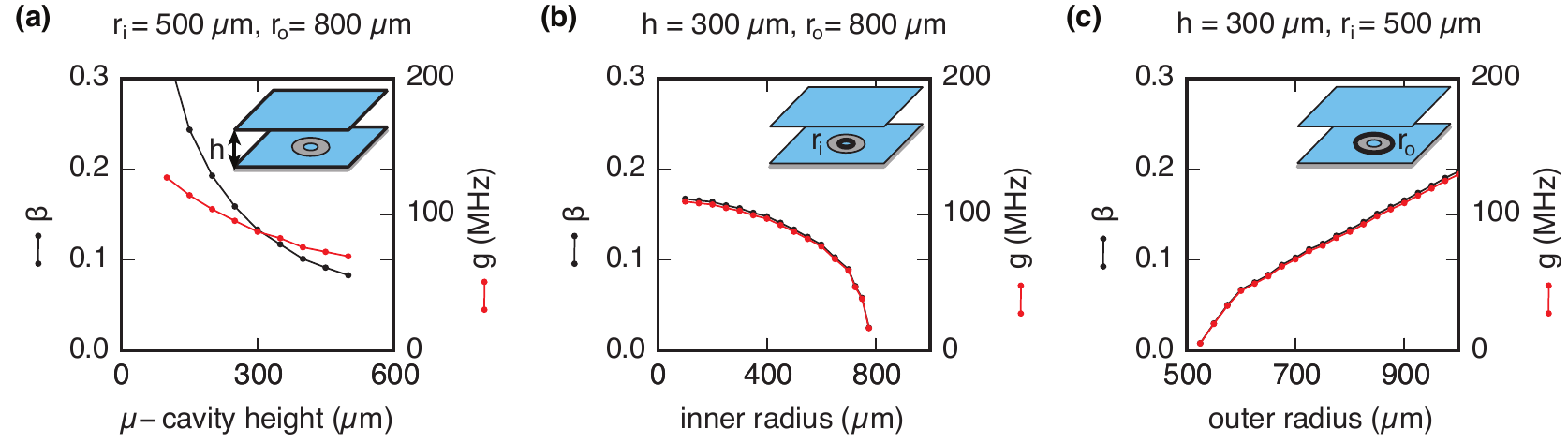}
\caption{
Relationships of the capacitance ratio $\beta$ (black) and coupling rate $g$ (red) versus three geometrical variations. These results were obtained using ANSYS\textsuperscript{\textregistered} Maxwell\textsuperscript{\textregistered} to model the capacitance matrix of a set of three conductors while including the silicon dielectric. The capacitance ratio $\beta= C_g/(C_g+C_p+C_j)$ is plotted in black and the coupling rate $g$ is plotted in red.  
     (a) Coupling increases with reduced micromachined cavity height by way of increasing gap capacitance $C_g$.
     (b) Coupling increases with smaller $r_i$, but with weakening dependence as $r_i$ becomes smaller than $h$.
     (c) Coupling increases with larger $r_o$, and the relationship is linear when $r_o>2r_i$. 
    }
\label{Fig:S1}
\end{minipage}
\end{figure*}

To get a rough estimate of the coupling, we can approach the problem in a similar way to the derivation found in Ref. \cite{Koch:2007}, which describes coupling of a dipole moment to the fields of a transmission line resonator.  First, we seek to obtain an expression for the dipole moment of the annulus structure in terms of the inner radius $r_i$ and outer radius $r_o$. 

Here we describe an analytic approach in two limiting cases. Begin with a large conductor plane with an annular ring removed, and assume that a voltage $V_0$ is applied across the annulus. There will there be an electric field across the gap as well as an electric field pattern out of the plane. The total field has dipole components pointing in opposite directions on either side of the plane. 

In the limiting case of an infinitely thin annulus, $r_i \approx r_o \equiv r$, the field is that of two opposing dipoles of identical magnitude pointing out of the annular plane in both directions. The magnitude of each is calculated in Ref. \cite{JacksonEM}: 
\begin{equation}
|\mathbf{p}|=\frac{4 \pi \epsilon_o}{3} V_0 r^2.
\end{equation}

In the limit of no central island at all ($r_i=0$), we have a circular aperture. If $r_o < \lambda$, we can make the following dipole approximation. In the presence of an incident E-field $E_o$, the field on the other side of the conducting plane can be approximated as a dipole with $r^3$ dependence, \cite{Collin}
\begin{equation}
|\mathbf{p}|=\frac{2}{3} \epsilon r_o^3 E_o.
\label{p_aperture}
\end{equation}

At the aperture of a resonant cavity, there is an oscillating dipole moment
$\mathbf{p}e^{i\omega t}$
which radiates power. Consider the average power radiated from a dipole into half-space: \cite{JacksonEM}
$$P_{\mathrm{rad}}=\frac{1}{2} \frac{1}{4\pi\epsilon_o} \frac{|\mathbf{p}|^2\omega^4}{3c^3}.$$

As the radius of the inner island goes from 0 to $r_o$, there must be some smooth correspondence between the fields in the two limits described above.  The actual system to which we want to apply this understanding is complicated by several effects. First is the presence of a nearby conductor: The distance from the annulus island to the wall of the micromachined cavity is on the same order as the annulus dimensions. Furthermore, the fields are affected by the presence of the dielectric substrate. Finally, we have eventually chosen a shape that differs from the circular annulus. In practice, an analytic expression for the fields or coupling strength is difficult and not the best avenue for design.

Instead, we do an electrostatic simulation of the relevant capacitances to investigate relationships to geometry.  Some such relationships are shown in Fig.~\ref{Fig:S1}. We plot the coupling rate $g=e V_0 \beta / \hbar$, which is a function of the capacitances, $\beta = C_g/(C_g+C_p+C_j),$ and $V_0 = \sqrt{\hbar \omega_{\mu} / 2 C}$, as they are defined in the main text. These trends are useful guides to design. Finally, we do a full driven modal finite element analysis simulation of the fields of the total design using ANSYS\textsuperscript{\textregistered} HFSS\textsuperscript{TM}.

\section{Device Fabrication}

The device is fabricated as three separate parts that are finally bonded together (see Fig.~2(a) of the main text). The process for the transmon chip begins with a $325~\mu$m thick (100)-orientation double-side-polished silicon wafer with resistivity $\rho > 10~$k$\Omega$cm. First, a gold pattern is defined on what will be one of the cavity walls using optical lithography and liftoff. The deposition is 10~nm Ti followed by 100~nm Au by e-beam evaporation. Second, the qubit is fabricated by e-beam lithography and a double-angle shadow e-beam evaporation of an aluminum Josephson junction, followed by liftoff. Third, the aluminum is masked with photoresist and the gold pattern on the wafer is electroplated with indium to a thickness of 10~$\mu$m. 

The fabrication process for the micromachined cavity chip is described in Ref. \cite{Brecht:2015apl}. It begins with a $1~$mm thick (100)-orientation silicon wafer with resistivity $\rho > 10~$k$\Omega$cm. A $22\times24~$mm rectangular pit is lithographically defined using a nitride mask. The pit is then etched to depth $300~\mu$m using a KOH etch bath at 85 \degree C. The concentration of KOH is 30 percent by weight in water and the etch bath also includes 1 percent isopropanol added as a surfactant. The wet etch is anisotropic with selectivity to silicon's $(100):(111)$ planes, resulting in a rectangular recess with sidewalls determined by the crystal planes and RMS surface roughness of $R_q$ = $20 - 40$ nm. The nitride mask is removed using BOE, which also removes salts left over from the etch bath. Then the wafer is coated in 100 nm evaporated gold before electroplating with indium to a thickness of $10~\mu$m.

Electroplating of the above described chips is done in an indium sulfamate plating bath solution purchased from Indium Corporation is operated at room temperature. In a wafer plating system purchased from Wafer Power Technology, a rotating Pt-Ti mesh serves as the insoluble anode and the wafer serves as the cathode onto which indium precipitates from the bath. A DC current of 300 mA is used to deposit indium at at rate of 150 nm/min to a final thickness of 10 $\mu$m on the full area of the 4 inch diameter substrate. The deposited material is known to be $>99.9 \%$ pure indium. 

The readout cavity is machined in OFHC copper and electroplated to a thickness of $30~\mu$m by an external vendor. The surface indium thickness is greater than that of the chip components in order to ensure complete coverage of the interior corners of the cavity.

The three components are bonded together between parallel plates of an Instron (5969) at 120\degree C in two steps. The components are sized such that the readout cavity completely surrounds the transmon chip and intersects the cavity chip. The two chips are bonded together first with 1 kN, followed by the bond of the readout cavity part to the cavity chip using 5 kN. For each bond, the force is ramped from 0 to the target over 1 minute and then held at target for 1 minute. Immediately prior to bonding, indium oxide is etched away with a solution of 10 percent hydrochloric acid in for 5 minutes, followed by DI water, acetone and methanol washing. (However, the transmon chip is not given a pre-bond etch because the hydrochloric acid damages the aluminum.) Once it is bonded, a cover piece made of copper (not pictured in Fig.~2) protects the stack and allows mounting to a thermalization bracket.

Once assembled, coaxial pin couplers to the readout cavity are tuned to provide $Q_{in}=430,000$ and $Q_{out}=42,000$. The device is thermally anchored to the baseplate of a dilution refrigerator reaching a base temperature of 10 mK. A cryoperm shield protects the device from magnetic fields. 

Note that in this assembly the readout cavity is larger than the micromachined cavity in all dimensions, and it is loaded with the bulk dielectric of the transmon chip. The readout is therefore lower frequency and a junction inductance $L_j$ is chosen such that the transmon frequency lies between those of the readout and micromachined cavities:  $\omega_r < \omega_q < \omega_\mu$.

\section{Transmon geometry}

With all of our design considerations in mind, we chose the following transmon geometry. The shape of the aperture cut from the outer conductor is a cardiod described by
\begin{equation}
\textup{outer } 
\left\{\begin{matrix}
x&=& \frac{r}{2}(2 \cos{t} - \cos{2t} + \frac{1}{2}) 
\\ 
y&=& \frac{r}{2}(2 \sin{t} - \sin{2t})
\end{matrix}\right.
, 0\le t < 2\pi.
\end{equation}
The inner island is a piriform curve described by
\begin{equation}
\textup{inner island } 
\left\{\begin{matrix}
x&=& b (\frac{1}{2}-2\sin{t})
\\ 
y&=& b \cos{t} (1+\sin{t})
\end{matrix}\right.
, 0\le t < 2\pi.
\end{equation}
These shapes were chosen in order to easily parameterize simulation variations. They may not necessarily be the optimal shapes for maximizing coupling strengths and minimizing surface losses. Nevertheless, these shapes produce a satisfactory design with parameter choices $r=750~\mu$m and $b=100~\mu$m. This shape is seen in Fig.~2(c) of the main text.

\section{Surface loss}
One design consideration is the effect of surface dielectric loss on qubit lifetime. The two-step simulation method described in Ref. \cite{Wang:2015} was performed on several representative design variations. For the chosen geometry, the simulated metal-substrate surface participation is $6.08\times10^{-4}$, substrate-air participation is $3.24\times10^{-4}$, and metal-air participation is $3.56\times10^{-5}$.

\section{Seam loss accounting}

We include in this supplementary note more detailed information on the In/Au/Al seam mentioned in the main text. We apply the model introduced in Ref. \cite{Brecht:2015apl}, which treats a seam as a distributed port around a path $\vec{l}$. Surface currents $\vec{J_s}$ that pass across the seam and dissipate power
\begin{equation}
P_{\mathrm{dis}} = \frac{1}{2G_{\mathrm{seam}}}
L \int_{\mathrm{seam}}|\vec{J_s} \times \hat{l}|^2 dl,
\end{equation}
where $L$ is the total length of the seam and $G_{\mathrm{seam}}$ is the total conductance. 

If it is damped solely by this power dissipation at the seam, a cavity mode of frequency $\omega$ and total energy $E_{\mathrm{tot}}$ has a quality factor $Q_i$ given by
\begin{equation}
\frac{1}{Q_i} = \frac{1}{\omega} \frac{P_{\mathrm{dis}}}{E_{\mathrm{tot}}} =
\frac{1}{G_{\mathrm{seam}}}
\left[\frac{L \int_{\mathrm{seam}} |\vec{J_s} \times \hat{l}|^2 dl}
{\omega \mu_o \int_{\mathrm{tot}} |\vec{H}|^2 dV} \right]
=\frac{y_{\mathrm{seam}}}{g_{\mathrm{seam}}},
\end{equation}
where the field $\vec{H}$ is integrated over the volume $V$ of the mode and $\mu_o$ is the magnetic permeability.  We identify the expression in square brackets as the admittance, $Y_{\mathrm{seam}}$, of the cavity presented to the seam. This admittance is zero when the seam is placed such that there are no perpendicular surface currents.

In order to compare intrinsic seam properties in different cavity constructions, assume uniform conductance and introduce a conductance per unit length $g_{\mathrm{seam}}=G_{\mathrm{seam}}/L$. Then we can isolate the conductance per unit length and identify the remainder of the expression as an admittance of the cavity as seen by the seam per unit length of the seam:
\begin{equation}
y_{\mathrm{seam}}= \frac{\int_{\mathrm{seam}} |\vec{J_s} \times \hat{l}|^2 dl}
{\omega \mu_o \int_{\mathrm{tot}} |\vec{H}|^2 dV}.
\end{equation}

Using this model, we can associate $y_{\mathrm{seam}}$ with the seam location and cavity fields and $g_{\mathrm{seam}}$ with materials properties in the seam region. In the next section, we analyze the placement of Al/Au/In seam in the device of the main text. Then, in the following section, we describe an experiment in which we have measured the conductance and conclude with some remarks on the materials.

\subsection{Admittance control}

\begin{figure}[h]
\linespread{1}
\centering
\includegraphics[scale = 1,angle=0]{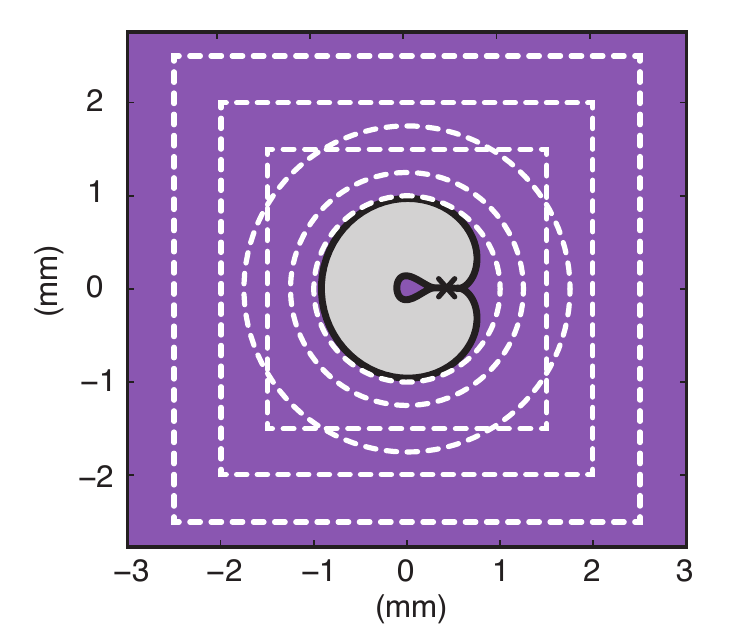}
\caption{
White dotted lines show seam paths that were considered for the In/Au/Al transition. The patchmon geometry (black lines) is fabricated in Al (purple) on Si (grey), defined by curves following equations 3 and 4 with $r=750~\mu$m and $b=100~\mu$m. The purple area inside the white dotted line is Al evaporated in a double-angle fashion simultaneous to the Josephson-junction. The purple area outside the white dotted line is In, and extends to form one wall of the micromachined cavity. This seam region is conceptually drawn in cross-section in Fig.~\ref{Fig:S3}. The corresponding seam admittances for both the qubit and cavity modes are numerically calculated in Table~\ref{seamtable}.
    }
\label{Fig:S2}
\end{figure}

We claim in the main text that one important design choice is the path of the seams. Several possible paths for the seam near the transmon are drawn as white dotted lines in Fig.~\ref{Fig:S2}.
The corresponding seam admittances for both the qubit and cavity modes are numerically calculated in Table~\ref{seamtable}. The table also calculates limits on qubit and cavity lifetimes using a conductance measured by an experiment described in the next section. We conclude that this seam may spoil the lifetime of the qubit or the cavity if it is not placed with intention to minimize its admittance. The path used for the actual device is the 3$\times$3 mm square.


\twocolumngrid

\vfill\break

\begin{table}[h]
\caption{
Limits on qubit and storage cavity lifetimes derived from seam admittances. The qubit mode $y_{\mathrm{seam}}$ values are found from HFSS simulation of the design featured in this work. Storage mode $y_{\mathrm{seam}}$ is calculated analytically by integrating surface currents of rectangular cavity's TE101 mode, and checked against simulation. The inferred lifetime limits are $T_1^{q,\mu} < g_{\mathrm{seam}}/y_{\mathrm{seam}}\omega_{q,\mu}$, assuming $g_{\mathrm{Al/Au/In}} = 4.2\times10^5~/\Omega$m, and using $\omega_q/2\pi = 7.3~$GHz, and $\omega_{\mu}/2\pi = 9.25~$GHz. 
*The last line computes limits imposed by the indium-to-indium bond around the perimeter of the micromachined cavity using $g_{\mathrm{In/In}} = 1\times10^8~/\Omega$m.
}
\label{seamtable}
\bgroup
\def\arraystretch{1.25}
\begin{tabular}{l|cccc}
						& $y_{\mathrm{seam}}^q$&$y_{\mathrm{seam}}^{\mu}$&max $T_1^q $&max $T_1^{\mu}$  \\
seam      & ($/\Omega$m) & ($/\Omega$m) & ($\mathrm{\mu s}$) & $(\mathrm{\mu s})$		\\ \hline \hline
In/Au/Al circle, r=1.00 mm	& 7.955 & 0.0114 & 1.1 & 640 \\ 
In/Au/Al circle, r=1.25 mm  & 1.565 & 0.0467 & 5.8 & 160 \\
In/Au/Al circle, r=1.75 mm  & 0.382 & 0.161 & 24 & 57 \\
In/Au/Al square, 3$\times$3 mm & 0.524 & 0.172 & 17 & 42 \\
In/Au/Al square, 4$\times$4 mm & 0.187 & 0.398 & 49 & 18 \\
In/Au/Al square, 5$\times$5 mm & 0.096 & 0.756 & 96 & 9.6 \\
In/In, cavity perimeter*	& 0.0239 & 15.96 & 91000 & 108  \\
\end{tabular}
\egroup
\end{table}

\begin{figure}[h]
\linespread{1}
\includegraphics[scale = 1,angle=0]{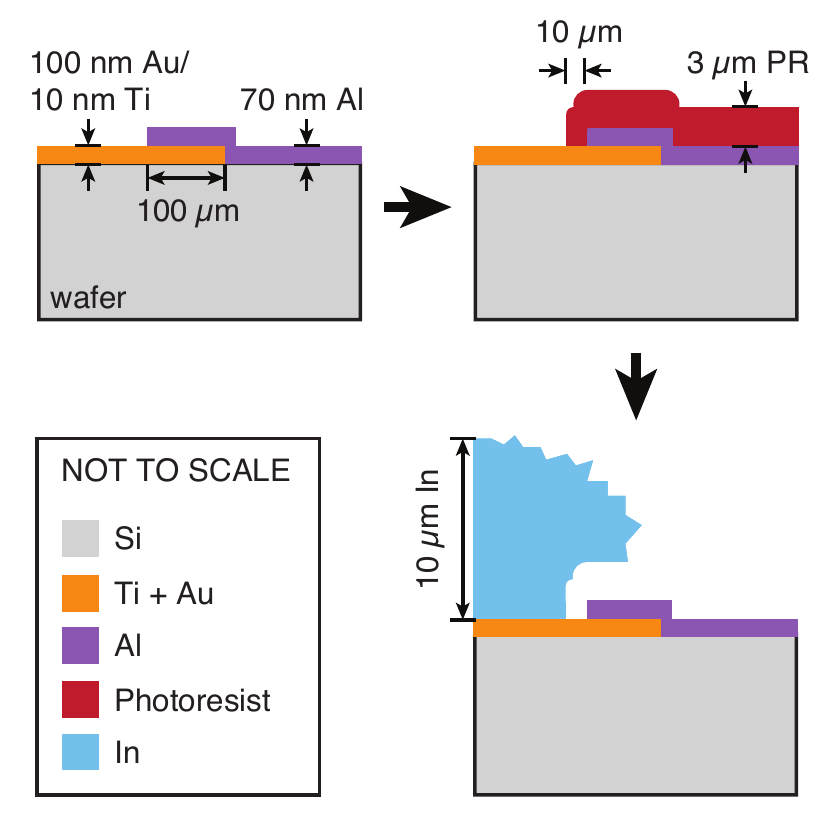}
\caption{
Conceptual drawings of the Al/Au/In seam region shown in cross-section during two intermediate fabrication steps (top) and the final product (bottom). The drawing portrays the rugged surface of the indium (blue). Note that the In overlaps the Al such that Au is not seen when imaged from above, as in Fig.~\ref{Fig:S4}. Nevertheless, current transiting this seam must pass from superconducting In through $100~\mathrm{\mu}$m normal Au to superconducting Al. 
    }
\label{Fig:S3}
\end{figure}

\FloatBarrier


\onecolumngrid

\subsection{Conductance measurement}

Now we take a closer look at the seam in question. Fig.~\ref{Fig:S3} shows the seam region in cross-section. Current transiting this seam must pass from superconducting In through $100~\mathrm{\mu}$m normal Au to superconducting Al.  Superconductor proximity effect is complicated by the possibility of intermetallics on both interfaces, which we discuss more later.

\begin{figure*}[!h]
\linespread{1}
\centering
\includegraphics[scale = 1,angle=0]{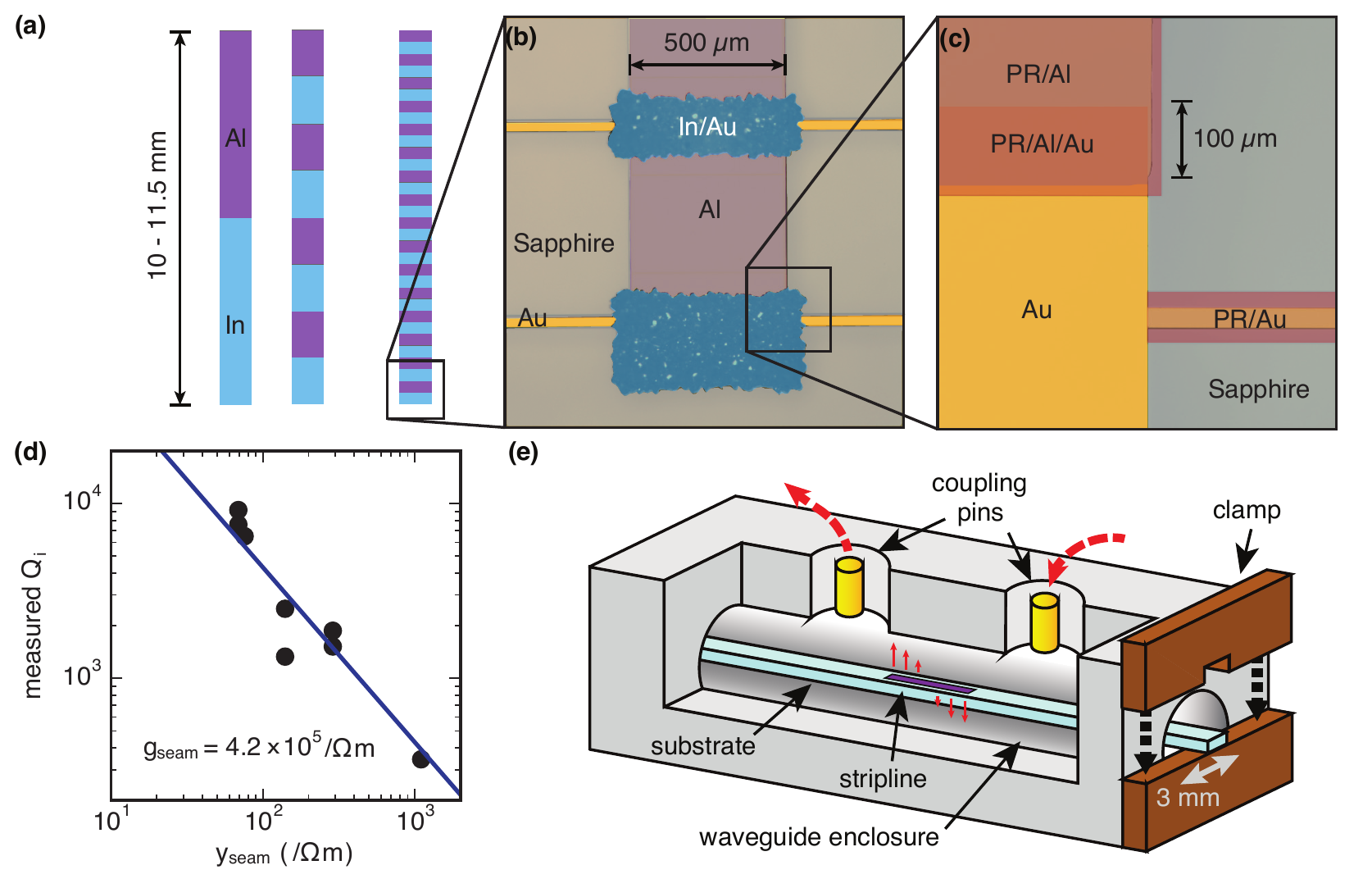}
\caption{Stripline resonators were used to measure the conductance of the seam in question. 
(a) Each device is a rectangular strip of conductor with transitions from Al (purple) to In (blue) that resemble those of the device featured in the main text as closely as possible. Devices were made with various numbers of transitions, and varying lengths to produce fundamental half-wave modes in the 7-8 GHz range.
(b) Optical microscope image of and end of one device. Three transition regions are visible. False coloring is applied for clarity.
(c) Optical microscope image showing an intermediate fabrication step, which is featured in cross section in Fig.~\ref{Fig:S3}. False coloring is applied for clarity.
(e) Measured internal quality factors of the several devices of varying seam admittances. The blue line is the best fit of the data to $Q_i=g_{\mathrm{seam}}/y_{\mathrm{seam}}$ using linear least-squares regression in the log-log domain, which yields $g_{\mathrm{seam}}=(4.2^{+0.6}_{-0.5})\times10^5~/\Omega$m.
(e) Each device is inserted into a tube-like package before measurement at 15 mK. Image used with permission from Ref. \cite{Axline:2016ue}.
    }
\label{Fig:S4}
\end{figure*}

The conductance of this seam, $g_{\mathrm{In/Au/Al}}$, was determined by a separate set of devices and an experiment that is illustrated in Fig.~\ref{Fig:S4}. In this experiment, we measured quality factors of stripline resonators in which the seam in question is the dominant loss mechanism. Microscrip resonators were fabricated on sapphire substrate with the same series of fabrication steps described for the transmon wafer in the main text. The resonator chips were loaded in a multiplexed version of the coaxial stripline package described in Ref. \cite{Axline:2016ue} (Fig.~\ref{Fig:S4}(e)) and measured in a cryogen-free dilution refrigerator at 15 mK. The measured internal quality factors of these resonators are plotted against seam admittance in Fig.~\ref{Fig:S4}(d). The blue line is the best fit of the data to $Q_i=g_{\mathrm{seam}}/y_{\mathrm{seam}}$ using linear least-squares regression in the log-log domain, which yields $g_{\mathrm{seam}}=(4.2^{+0.6}_{-0.5})\times10^5~/\Omega$m. We expect that improvement is possible using cleaning methods, the exploration of which can be the subject of future study. 


\subsection{Conductance discussion}

The presence of Au is certainly non-ideal. It exists in this process to serve as the conducting layer onto which In is electroplated. Most other conductors are not suitable for In electroplating because of either oxides or incompatible electronegativity. Cu is another appropriate metal for In electroplating. However, we find that DC contact resistance for Cu/Al is lower than that of Au/Al (at least without experimenting with cleaning procedures before Al deposition). Therefore, Au is used as the electroplating seed layer. Furthermore, all Al on the wafer must be completely masked during electroplating  to prevent its destruction.

It is worth noting that Au is known to interact with the nearby metals in worrisome ways. In particular, Au will form intermetallic compounds with both Al and In. 
$\mathrm{Au_x Al_y}$ intermetallics are known problems in microelectronics and wire-bonding. Two such legendary intermetallics are ``white plague'' ($\mathrm{Au_5 Al_2}$), which has low conductivity, and ``purple plague'' ($\mathrm{Au Al_2}$), which leads to voids at the interface. Our process avoids the high temperatures ($>624\degree$C) that produce these compounds, but it is possible that diffusion takes place to form $\mathrm{Au Al_2}$ (occurring at $>400\degree$C, possibly occurring slowly at lower temperatures), which is also a poor conductor.

Au also diffuses into In, and the intermetallic $\mathrm{AuIn}_2$ forms when the temperature is elevated above the melting point of In: $156\degree$C. This intermetallic is a known problem in indium bump manufacturing methods that involve reflow of the indium\cite{Huang2010}, and diffusion rates of have been measured by Refs. \cite{Huang2010,InAu2}. Our process does not involve molten In. There is concern, that any contaminants in the In, Au or other, can reduce its ductility, which has negative consequences for the conductance of In bonds between layers. We do not know how this may affect the superconductivity of In. 

Though evidence of these intermetallics has not been discovered in our samples, they are considered cause for concern. Therefore, fabrication procedures that eliminate the Au layer altogether are being explored for future use. In particular, In can be deposited via thermal evaporation instead of electroplating. Alternatively, a thin Ti layer between Au and In could be incorporated into our process, as it is an effective diffusion barrier at low temperature in Ref.~\cite{InAu2}.




%

\end{document}